\documentclass[preprint,nofootinbib,aps,superscriptaddress,eqsecnum]{revtex4-1} %\documentclass[11pt, a4paper]{article}
 \pdfoutput=1
 \usepackage{graphicx}
 \usepackage{amsmath}
 \usepackage{physics}
\usepackage{amsfonts}
\usepackage{amssymb}
\usepackage{hyperref}
\usepackage{gensymb}
\usepackage[T1]{fontenc}
\usepackage{array}
\usepackage{booktabs}
\usepackage{caption}
\usepackage{subcaption}
\usepackage{changepage}
\usepackage[dvipsnames]{xcolor}
\usepackage{changes}
\usepackage{placeins}
\usepackage{multirow}

\def\bea{\begin{eqnarray}}
\def\eea{\end{eqnarray}}
\def\be{\begin{equation}}
\def\ee{\end{equation}}

\begin{document}
  
\title{Exploring the nature of neutrinos in a dissipative environment.}
 
\author{Chinmay Bera}
\email[Email Address: ]{chinmay20pphy014@mahindrauniversity.edu.in}
\affiliation{Department of Physics, \'Ecole Centrale School of Engineering - Mahindra University, Hyderabad, Telangana, 500043, India}

\author{K. N. Deepthi}
\email[Email Address: ]{nagadeepthi.kuchibhatla@mahindrauniversity.edu.in}
\affiliation{Department of Physics, \'Ecole Centrale School of Engineering - Mahindra University, Hyderabad, Telangana, 500043, India}
\affiliation{Indiana University, Bloomington, Indiana 47405, USA}

\begin{abstract}
In this study, we explore the scope of determining the neutrino nature in long-baseline neutrino oscillation experiments considering the effect of environmental decoherence in neutrino evolution. Assuming an open quantum system framework, we numerically analyze the two flavor neutrino oscillation probabilities. We observe that the transition probabilities accommodate the Majorana phase in the presence of dissipative environment. Considering this phenomenology, we study the effect of Majorana phase on these probabilities and investigate the sensitivity of T2K, ESSnuSB, NOvA, T2HKK and DUNE to differentiate between Dirac and Majorana neutrinos. 
\end{abstract}

\pacs{}
\maketitle

%%%%%%%%%%%%%%%%%%%%%%%%%%%%%%%%%%%%%
\section{Introduction}\label{introduction}
%%%%%%%%%%%%%%%%%%%%%%%%%%%%%%%%%%%%%%
One of the unresolved puzzles in neutrino physics is whether neutrinos are Dirac or Majorana fermions. 
In the Dirac description, neutrinos are different from their antiparticles and the lepton number (L) is conserved. 
Whereas in the Majorana picture, neutrinos and anti-neutrinos are not physically distinguishable. If neutrinos are Majorana fermions, neutrinoless double beta decay ($0\nu 2 \beta$) process ($X^Z_A \rightarrow Y^A_{Z + 2} + 2 e^-$) could occur with the violation of lepton number ($\Delta L = 2$)~\cite{Vergados:2012xy,Bilenky:2014uka}. To this day, there is no experimental evidence for $0\nu 2 \beta$ process. %Several running and upcoming experiments are looking for this event. 
The most recent combined analysis of KamLAND-Zen 400 (2011-2015) and KamLAND-Zen 800 (started 2019) data predicts the half-life of $T_{0\nu 2 \beta}^{1/2}>2.3 \times 10^{26}$ yr at 90$\%$ C.L. and this corresponds to an effective neutrino mass ($m_{\beta \beta}$) less than $36-156$ $meV$~\cite{KamLAND-Zen:2022tow}. The smallness of neutrino mass has been successfully explained in beyond the standard model (BSM) physics, where neutrinos are assumed to be Majorana particles. Therefore, to resolve this puzzle, holds a strong theoretical motivation for BSM physics.

Experimental data from nearly two decades has established neutrino oscillations as a leading mechanism for neutrino flavour transitions. However, sub-leading effects like non-standard interactions, quantum decoherence, neutrino decay, still hold a torch for new physics scenarios beyond the standard model. Therefore it is the need of the hour to investigate the implications of these sub-leading effects. 

In this work, we study the effect of quantum decoherence on the neutrino oscillation probabilities in an open quantum system framework. The open quantum system is modelled by considering the interaction of the neutrino subsystem with the environment~\cite{Farzan:2008zv,oliveira2010quantum,Bakhti:2015dca,Guzzo:2014jbp,BalieiroGomes:2016ykp,Capolupo:2018hrp}. The interactions of this kind could originate from the effects of quantum gravity \cite{Hawking_Unpredictability, Hawking:1995ag, Hawking_Wormholes, Hawking:1998jf}, strings and branes  \cite{Ellis:1992eh, Ellis:1992pm, Benatti:2000wu} at the Plank scale. Consequently, they manifest as dissipation effects and modify the neutrino oscillation probabilities. Moreover, in ref.~\cite{Benatti:2001fa,oliveira2010quantum,Richter:2017toa,Capolupo:2018hrp,Buoninfante:2020iyr,Carrasco-Martinez:2020mlg}, it has been shown that one can probe the nature of neutrinos when the neutrino system interacts with the environment.

In the present work, we analyse the effect of a dissipative environment on the neutrino oscillation probabilities in a two neutrino framework including the matter effect. The probabilities depend on the Majorana phase and thus provide a window to probe the nature of neutrinos for different baselines. In this context, we study the transition probabilities at different baselines of T2K (295 km), NOvA (810 km) and the upcoming experiments ESSnuSB (540 km), T2HKK (1100 km), DUNE (1300 km) experiments and their dependency on the Majorana phase. In addition, we present the relative events rate (difference in Dirac and Majorana events) for these experiments and obtain the sensitivities to the Majorana phase.

This paper is structured as follows:  In section~\ref{sec:mathematical-formulation}, we present a basic formalism to determine the neutrino oscillation probabilities in matter while assuming decoherence. We discuss in section~\ref{sec:simulation-details}, the experimental simulation details of the neutrino oscillation experiments that have been used in this study. In section~\ref{sec:numerical-analysis}, we show the oscillation probabilities relevant to these experiments and discuss their implications by considering four feasible combinations of decoherence parameters. In section~\ref{sec:event-sensitivity}, we present the event rates w.r.t neutrino energy and sensitivity for all the experiments considered \footnote{barring T2K results, the reason for this is presented in the beginning of the section~\ref{sec:numerical-analysis}.}. We finally summaries our findings in section~\ref{sec:conclusion}.

%%%%%%%%%%%%%%%%%%%%%%%%%%%%%%%%%%%%%%%%%%%%%%%%%%
\section{Formalism of neutrino oscillations in matter assuming decoherence}\label{sec:mathematical-formulation}
%%%%%%%%%%%%%%%%%%%%%%%%%%%%%%%%%%%%%%%%%%%%%%%%%%
In an open quantum system framework, the neutrino subsystem interacts weakly with the environment leading to a loss of coherence in the subsystem. 
The decoherence phenomenon in Markovian systems is described by the Lindblad-Kossakowski master equation~\cite{Lindblad:1976g,GKS:1976vit}
\begin{equation}\label{lindblad}
    \mathcal{L} \rho(t) = \frac{\partial\rho}{\partial t} = -i[H_{eff} , \rho(t)] + \mathcal{D}[\rho(t)]~.
 \end{equation} 
 Here, the infinitesimal generator $\mathcal{L}$ and its action on the density matrix $\rho(t)$ depends on the effective Hamiltonian $H_{eff}$ and the dissipative term $\mathcal{D}$.
The dissipative factor $\mathcal{D}$ has the following form~\cite{gorini1978properties},
\begin{equation}\label{dissipative}
    \mathcal{D}[\rho] = - \frac{1}{2}\sum_{j=0}^{N^2 - 1}(A_j^\dagger A_j \rho + \rho A_j^\dagger A_j) + \sum_{j=0}^{N^2 - 1} A_j \rho A_j^\dagger~.
\end{equation}
Here $A_j$ are the operators that depend on the dimensions of the system. In an N-level system, $A_j$ has $N \times N$ dimension and they form $N^2 - 1$ linearly independent basis without including identity matrix. For the two flavor mixing, $A_j$ are the Pauli matrices $\sigma_i$ and that for the three flavor are represented by Gell-Mann matrices. In the former case the density matrix $\rho$ and operator $A_j$ in eq.~(\ref{dissipative}), can be written as, $\rho = \frac{1}{2}\rho_\mu \sigma_\mu$ , $A_j = a_{\mu,j}\sigma_\mu$ where, $\mu = 0, 1, 2, 3$, $\sigma_0$ is $2 \times 2$ identity matrix.

In this work, we consider ultra-relativistic electron neutrinos ($\nu_e$) and muon neutrinos ($\nu_{\mu}$) in two dimensional Hilbert space. The flavor states ($\nu_e$, $\nu_{\mu}$) are related to the mass states ($\nu_1$, $\nu_2$) by a $2 \times 2$ unitary mixing matrix  
\begin{equation}\label{mixing}
    \begin{pmatrix}
    \nu_e\\
    \nu_\mu
    \end{pmatrix} 
    = \begin{pmatrix}
    \cos\theta  & \sin\theta~e^{i \phi}\\
    - \sin\theta & \cos\theta~e^{i \phi}
    \end{pmatrix}
    \begin{pmatrix}
    \nu _1\\
    \nu_2
    \end{pmatrix},
    \end{equation}
    
where $\theta$ is the mixing angle and $\phi$ is the Majorana phase.

 In eq.~(\ref{lindblad}), one can note that the evolution of $\rho$ depends on a time independent effective Hamiltonian  $H_{eff}$, which is combination of vacuum Hamiltonian ($H_{vac}$) and matter interaction Hamiltonian ($H_{mat}$) 
\begin{align}\label{H_eff}
    H_{eff} = \begin{pmatrix}
   - \omega & 0\\
   0 &  \omega
   \end{pmatrix} + A
    \begin{pmatrix}
    \cos^2\theta  & \frac{1}{2} \sin2\theta  e^{-i \phi}\\
     \frac{1}{2} \sin2\theta  e^{i \phi} & \sin^2\theta
    \end{pmatrix} = h_\mu \sigma_\mu~ ,
\end{align}
where, in the first matrix, $\omega = \frac{\Delta m^2}{4E}$ contains the square mass difference of two mass eigen-states ($\Delta m^2$) and $E$ represents the neutrino energy.  The interaction of the neutrinos with the matter is given by the second term (interaction Hamiltonian), where $A = \sqrt{2}G_Fn_e$. Here $G_F$ is called the Fermi constant and the electron number density in the medium is denoted by $n_e$. The effective Hamiltonian in the SU(2) representation, takes the form, $H_{eff} = h_\mu \sigma_\mu$, where $h_\mu$ are the coefficients of the Pauli matrices (generators in SU(2) representation).

The dissipative matrix $\mathcal{D}_{mn}$ (based on positivity and trace-preserving conditions) depends on six real and independent parameters $\alpha,~\beta,~\gamma,~a,~b,~c$. The matrix $\mathcal{D}_{mn}$ is given by,
    
   \begin{align} \label{Dmn}
    \mathcal{D}_{mn} = -2\begin{pmatrix}
    0&0&0&0\\
    0&\alpha&\beta&\gamma\\
    0&\beta&a&b\\
    0&\gamma&b&c
    \end{pmatrix}~,
\end{align}
where $m,n=0,1,2,3$. 

Now, the time evolution of the density matrix $\rho$ in eq.~(\ref{lindblad}) can be expressed through Schr$\Ddot{o}$dinger like equation 
\begin{eqnarray}\label{sch}
\frac{d}{dt}\ket{\rho(t)} = -2 \mathcal{H}\ket{\rho(t)}~,
\end{eqnarray}
Here, $\mathcal{H}$ can be obtained by substituting eqs.~(\ref{dissipative}), (\ref{H_eff}), (\ref{Dmn}) in eq.~(\ref{lindblad}) as 
    \begin{align}\label{H_sch}
    \mathcal{H} = \begin{pmatrix}
    0 & 0 & 0 & 0\\
    0 & \alpha & \beta + \xi & \gamma - \lambda \sin\phi\\
    0 & \beta - \xi & a & b + \lambda \cos\phi\\
    0 & \gamma + \lambda \sin\phi & b - \lambda \cos\phi & c
    \end{pmatrix}
\end{align}

and \begin{align}
    \xi = \frac{A}{2}\cos2\theta - \omega, \hspace{0.5cm} \lambda = \frac{A}{2}\sin2\theta
\end{align}

After imposing trace preserving condition $\Dot{\rho_0}(t) = 0$, eq.~(\ref{sch}) will take the following form
\begin{align}
    \begin{pmatrix}
     \Dot{\rho_1}(t) \\ \Dot{\rho_2}(t) \\ \Dot{\rho_3}(t)   
    \end{pmatrix} = -2 \mathcal{H} \begin{pmatrix}
     {\rho_1}(t) \\ {\rho_2}(t) \\ {\rho_3}(t)
    \end{pmatrix}.
\end{align}
Since $\Dot{\rho}_0(t) = 0$, we take a reduced form of $3 \times 3$ $\mathcal{H}$ matrix from eq.~(\ref{H_sch}).
Hence, the evolved state at time t is written as

\begin{align}\label{evolveddensity}
    \begin{pmatrix}
     \rho_1(t) \\ \rho_2(t) \\ \rho_3(t)   
    \end{pmatrix} = \mathcal{M}(t) \begin{pmatrix}
     {\rho_1(0)} \\ {\rho_2(0)} \\ {\rho_3(0)}
    \end{pmatrix} 
\end{align}
where 
\begin{align}\label{basline-dependency}
    \mathcal{M}(t) = e^{-2\mathcal{H}t} = \mathcal{S}e^{-2\mathcal{H'}t}\mathcal{S}^{-1}
\end{align}
is a $3 \times 3$ matrix.
Here, $\mathcal{S}$ is the similarity transformation matrix and $\mathcal{H'}$ is the eigenvalue matrix of $\mathcal{H}$.

The density matrix at arbitrary time t is
\begin{align}\label{density}
 \rho(t) = \frac{1}{2}
    \begin{pmatrix}
    \rho_0(t) + \rho_3(t) & \rho_1(t) - i \rho_2(t)\\
    \rho_1(t) + i \rho_2(t) & \rho_0(t) - \rho_3(t)
    \end{pmatrix}~.
\end{align}

Using eq.~(\ref{mixing}), the density matrix at initial time t = 0 can be obtained as
\begin{align}
    \rho_{\nu_e}(0) = \begin{pmatrix}
    \cos^2\theta & \frac{1}{2}\sin2\theta~e^{-i\phi}\\
    \frac{1}{2}\sin2\theta~e^{i\phi}
& \sin^2\theta
\end{pmatrix}
\end{align} 
and
\begin{align}\label{muondensity}
    \rho_{\nu_\mu}(0) = \begin{pmatrix}
    \sin^2\theta & - \frac{1}{2}\sin2\theta~e^{-i\phi}\\
    - \frac{1}{2}\sin2\theta~e^{i\phi}
& \cos^2\theta
\end{pmatrix}~.
\end{align}

Further, the evolved state $\rho_{\nu_\mu}(t)$ can be obtained using eq.~(\ref{evolveddensity})-(\ref{muondensity}). The probability of transition of an initial state $\nu_\alpha$ to a final state $\nu_\beta$ can be evaluated using 
\begin{align}
    P_{\alpha\beta}(t) = Tr[\rho_{\nu_\alpha}(t) \rho_{\nu_\beta}(0)]~.
\end{align}
Upon substitution, the appearance $ P_{\mu e}(t)$ and disappearance $P_{\mu\mu}(t)$ probabilities are obtained to be 
 \begin{equation}\label{nue-app prob}
 \begin{aligned}
     & P_{\mu e}(t) = \\
     & \frac{1}{2} \Bigl[1 - M_{33} \cos^2(2\theta) - \sin(2 \theta) \cos(2\theta) \bigl\{(M_{23} + M_{32}) \sin\phi\\
       & + (M_{13} + M_{31})\cos\phi \bigr\} - \sin^2(2\theta) \bigl\{(M_{12} + M_{21})\sin\phi \cos\phi\\ 
       & + (M_{11}\cos^2\phi + M_{22}\sin^2\phi)\bigr\}\Bigr]~,
 \end{aligned}
 \end{equation}
 %%%%%%%%%%%%%%%%%%%%%%%%%%%%%%%
  \begin{equation}\label{numu-dis prob}
 \begin{aligned}
    & P_{\mu\mu}(t) = \\
    & \frac{1}{2} \Bigl[1 + M_{33} \cos^2(2\theta) + \sin(2 \theta) \cos(2\theta) \bigl\{(M_{23} + M_{32}) \sin\phi\\
      & + (M_{13} + M_{31})\cos\phi \bigr\} + \sin^2(2\theta) \bigl\{(M_{12} + M_{21})\sin\phi \cos\phi\\
      & + (M_{11}\cos^2\phi + M_{22}\sin^2\phi)\bigr\}\Bigr] = 1 -  P_{\nu_\mu \rightarrow \nu_e}(t)~,
 \end{aligned}
 \end{equation}
%%%%%%%%%%%%%%%%%%%%%%%%%%%%% 
where $M_{ij}$ are the matrix elements of $\mathcal{M}$ and $i,j = 1,2,3$.

The probability of antineutrinos under the same conditions can be obtained from the probability of neutrinos by replacing the $A \rightarrow -A$ and $U \rightarrow U^*$ in the calculation. Then $P_{\bar{\mu}\bar{e}}(t)$ and $P_{\bar{\mu}\bar{\mu}}(t)$ are found to be

\begin{equation}\label{antinue app prob}
 \begin{split}
     & P_{\bar{\mu}\bar{e}}(t) = \\
     & \frac{1}{2} \Bigl[1 - M_{33} \cos^2(2\theta) - \sin(2 \theta) \cos(2\theta) \bigl\{(M_{13} + M_{31})\cos\phi \\
       & - (M_{23} + M_{32}) \sin\phi\bigr\} + \sin^2(2\theta) \bigl\{(M_{12} + M_{21})\sin\phi \cos\phi\\ 
       & - (M_{11}\cos^2\phi + M_{22}\sin^2\phi)\bigr\}\Bigr]~,
 \end{split}
 \end{equation}

 and 

 \begin{equation}\label{antinumu dis prob}
 \begin{split}
     P_{\bar{\mu}\bar{\mu}}(t)  = 1 -  P_{\bar{\mu} \bar{e}}(t)
 \end{split}~.
 \end{equation}

The eqs.(\ref{nue-app prob} - \ref{antinumu dis prob}) show that in the presence of decoherence the neutrino and anti-neutrino oscillation probabilities  depend on the Majorana phase $\phi$. This modified oscillation probabilities open a window to explore the nature of neutrinos in the current and the upcoming neutrino oscillation experiments. In this context, we analyse the probabilities of various long baseline neutrino oscillation experiments and their dependency on the Majorana phase $\phi$ in the presence of decoherence.  We consider $t=L$ in the natural units, where L is the distance traveled by the neutrino beam. In the following section, we give a brief account of the five experiments considered in this study. 
%%%%%%%%%%%%%%%%%%%%%%%%%%%%%%%%%%%%

%%%%%%%%%%%%%%%%%%%%%%%%%%%%%%%%%%%%%%%%%
\section{Experimental details}\label{sec:simulation-details}
%%%%%%%%%%%%%%%%%%%%%%%%%%%%%%%%%%%%%%%%%%%%%%%%%%%%
\begin{table*}[htb!]
    %\centering
    \caption{Experimental simulation details.}
    \label{table:exp_details}
    \begin{tabular*}{\textwidth}{@{\extracolsep{\fill}}lrrrrl@{}}
\hline
    \textbf{Experiment} & \textbf{T2K} & \textbf{T2HKK} & \textbf{ESSnuSB} & \textbf{NO$\nu$A} & \textbf{DUNE} \\
    \hline
    %Location & Japan & Japan & Lund, Europe & USA & USA \\
    Status & Operating & Proposed & Proposed & Operating & Construction \\
    Beam power & 200 kW & 1.3 MW & 5 MW & 700 kW & 1.2 MW \\
    %Exposure & & & $2.5 \times 10^{23}$ POT/year & & \\
    Run time & 3($\nu$) + 3($\bar{\nu}$) & 5($\nu$) + 5($\bar{\nu}$) & 5($\nu$) + 5($\bar{\nu}$) & 6($\nu$) + 3($\bar{\nu}$) & 5($\nu$) + 5($\bar{\nu}$) \\
    Baseline (L) & 295 km & 1100 km & 540 km & 810 km & 1300 km \\
    Flux peak & 0.6 GeV & 0.7 GeV & 0.2 GeV & 1.8 GeV & 2.8 GeV \\
    Density ($\rho$) & 2.8 $gm/cm^3$ & 2.88 $gm/cm^3$ & 2.8 $gm/cm^3$ & 2.84 $gm/cm^3$ & 2.848 $gm/cm^3$ \\
    Off-axis angle & 2.5\degree & 1.5\degree & 0\degree & 0.8\degree & 0\degree \\
    detector mass & 22.5 kt & 187 kt & 500 kt & 14 kt & 40 kt \\
    Target material & Pure water & Pure water & Pure water & LS & LiArTPC \\
    %$\nu_e$ events/year & & & & & \\
    %$\bar{\nu}_e$ events/year & & & & & \\
    Refs. & \cite{abe2013t2k} & \cite{hyper2018physics} & \cite{baussan2012use} & \cite{NOvA:2021nfi} & \cite{DUNE:2021cuw} \\
    \hline
    \end{tabular*}
\end{table*}
%%%%%%%%%%%%%%%%%%%%%%%%%%%%%%%%%%%%%%
%%%%%%%%%%%%%%%%%%%%%%%%%%%%%%%%%%%%%%%%%%%%%%%%%%%%%%%%%%%%%%%%%%%%%%%%%%

T2K and T2HKK : Tokai to Kamiokande (T2K) experiment ~~\cite{abe2013t2k} is an off-axis (off-axis angle (OAA) of $2.5^{\degree}$) oscillation experiment with Japan Proton Accelerator Research Complex (J-PARC) based $\nu_{\mu}$ beam facility. The far detector of volume 22.5 kt is placed at Kamiokande with a baseline of 295 km. The motivation of this experiment is to precisely measure oscillation parameters, $\theta_{13}$, $\theta_{23}$ and $\Delta m^2_{32}$.
The T2HKK experiment ~~\cite{hyper2018physics} is proposed to have an off-axis $\nu_{\mu}$ beam (OAA ranging from 1-3$\degree$) from J-PARC facility, to travel a distance of 1100 km before it reaches the Water Cherenkov detector of 187 kt based in Korea. 

NOvA : The NuMI Off-axis $\nu_e$ Appearance Experiment ~~\cite{NOvA:2021nfi} is an ongoing neutrino oscillation experiment with a baseline of 810 km. It has an off-axis (OAA 0.8$\degree$) muon neutrino beam with a peak energy of $\sim$2 GeV. NOvA has a near detector at the Fermilab site and NuMI beam focused towards a far detector of volume 14 kt, placed at Minnesota. The main goal is to understand the atmospheric neutrino flavor transition and also measure atmospheric mass square difference to the higher precision level ($10^{-4}$ $eV^2$). 

ESSnuSB : The European Spallation Source Neutrino Super Beam (ESSnuSB) experiment~\cite{baussan2012use} is an upcoming oscillation experiment with a baseline of 540 km. The peak energy of the neutrino beam is around 0.2 GeV which is the energy corresponding to the second oscillation maxima. A 500 kt Water Cherenkov (WC) far detector is placed underground in ESS site at Lund. This experiment is sensitive to observe leptonic CP-violation phase at 5$\sigma$ confidence level.

DUNE : The Deep Underground Neutrino Experiment (DUNE)~\cite{DUNE:2021cuw} is an on-axis accelerator based long baseline neutrino experiment. The DUNE mainly consists of a beamline, a near detector at Fermilab and a far detector at Sanford Underground Research Facility (SURF) which is 1300 km away from the near detector. It consistently will measure the neutrino events having broad range of energy ($1-8$ GeV). The neutrino flux is peaked around 2.8 GeV corresponding to the energy at the first oscillation maxima. This experiment will help to determine charge-parity (CP) violation phase and the neutrino mass ordering with very high precision.

\begin{table}[!htb]
\centering
\caption{Standard neutrino oscillation parameters are taken from NuFIT 5.2~\cite{Esteban:2020cvm} and decoherence parameters are taken from ref.~\cite{DeRomeri:2023dht}.}
     \label{table:osc-deco-params}
       \begin{tabular*}{\columnwidth}{@{\extracolsep{\fill}}ll@{}}
        \hline
        \textbf{Parameter} & \textbf{Value}  \\
        \hline
        $\Delta m^2_{32}$ & 2.51 $\times 10^{-3}$ $eV^2$  \\
        $\sin^2\theta_{13}$ & 0.02205  \\
        $\alpha$ = $a$  & 2.2 $\times 10^{-23}$ GeV \\
        $c$ & 3.2 $\times 10^{-23}$ GeV \\
        $\beta$ & 0.61 $\times 10^{-23}$ GeV  \\
        $b$ = $\gamma$ & 0.53 $\times 10^{-23}$ GeV \\
        \hline
      \end{tabular*}
\end{table}

A few important specifications of each of the above experiments have been tabulated in table~\ref{table:exp_details} to facilitate an ease in comparison. The details of the oscillation parameters and the decoherence parameters considered in this work are listed in table~\ref{table:osc-deco-params}.
%%%%%%%%%%%%%%%%%%%%%%%%%%%%%%%%%%%%%%%%%%%%%%%%
%%%%%%%%%%%%%%%%%%%%%%%%%%%%%%%%%%%%%%%%%%%%%%
\section{Numerical Analysis}\label{sec:numerical-analysis}
%%%%%%%%%%%%%%%%%%%%%%%%%%%%%
 We consider the two-flavor neutrino oscillation analysis of $\nu_{\mu}$-$\nu_e$ channel in a dissipative medium. We use the formalism presented in section~\ref{sec:mathematical-formulation}. 
 We provide the probability versus neutrino energy plots corresponding to the eqs.~(\ref{nue-app prob} - \ref{antinumu dis prob}), as these are the channels to be studied in the experiments T2K, NOvA, ESSnuSB, T2HKK and DUNE. We use GLoBES~\cite{Huber:2004ka,Huber:2007ji} software packages to execute all the simulations in this work. We incorporate a new probability engine in GLoBES to implement decoherence.
 
 Firstly, for simplicity,  we assume all the off-diagonal elements of the decoherence matrix in eq.~(\ref{Dmn}) to be zero. Secondly, we assume non-zero diagonal and one non-zero off-diagonal elements. Finally, for completeness we also present the case where all the elements of $\mathcal{D}_{mn}$ are non-zero. We assume the neutrino mass ordering as normal ordering throughout the paper unless otherwise mentioned.
 %%%%%%%%%%%%%%%%%%%%%%%%%%%%%%%%%%%%%%%%%%%%
 %%%%%%%%%%%%%%%%%%%%%%%%%%%% Pmue_E 

\subsection{Non-zero diagonal elements in $\mathcal{D}_{mn}$:}\label{subsec:non-zero-diagonal}
%%%%%%%%%%%%%%%%%%%%%%%%%%%%%%%%%%%%%%%%%%%%%%%%%%%%%%%%%%%%%%%%%%%%%%%%%%%%%%%%%%%%%%%%%%%%%%%%%%%%%%%%%%%%%%%%%%%%%%%%%%%%%%

\begin{figure*}[hbt!]
%\begin{center}
\centering
\includegraphics[width=0.45\linewidth]{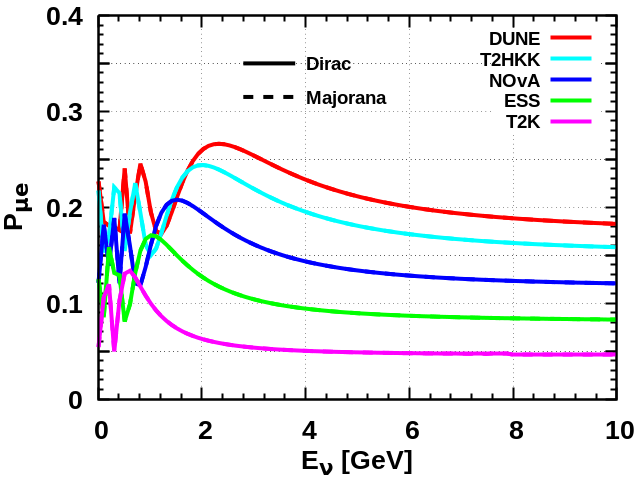}
\includegraphics[width=0.45\linewidth]{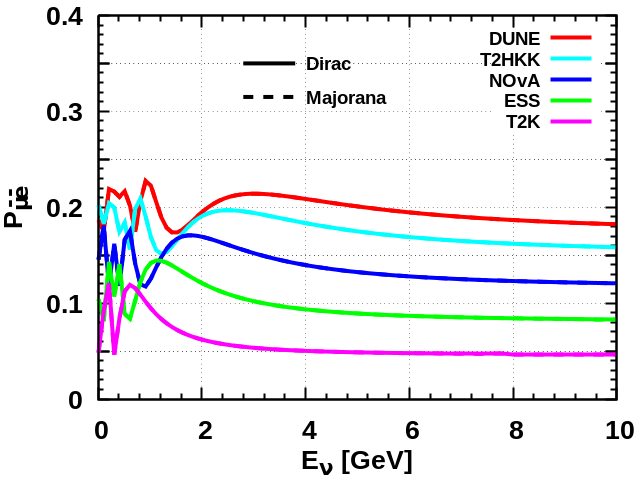}
\caption{\footnotesize $\nu_e$-appearance (left) and $\bar{\nu}_e$-appearance (right) probabilities with respect to energy E (assuming NH). The magenta, green, blue, cyan and red curves correspond to the baselines of 295 km, 540 km, 810 km, 1100 km and 1300 km respectively. Approximations: (a) only diagonal elements in $\mathcal{D}_{mn}$ are non-zero with $\phi = \frac{\pi}{4}$.}
\label{fig:prob-nue_diag}       
\end{figure*}

Firstly, we assume that the decoherence matrix $\mathcal{D}_{mn}$ in eq.~{\ref{Dmn}} has only non-zero diagonal elements and impose that the off-diagonal elements are zero. Under this assumption $\mathcal{D}_{mn}$ in eq.~{\ref{Dmn}} takes a simple form 
   \begin{align} \label{diag_Dmn}
    \mathcal{D}_{mn} = -2\begin{pmatrix}
    0&0&0&0\\
    0&\alpha&0&0\\
    0&0&a&0\\
    0&0&0&c
    \end{pmatrix}~,
\end{align}
where we assume, $\alpha = a \neq c$. Now, using the density matrix formalism presented in sec.II, we numerically obtain the oscillation probabilities $P_{\alpha\beta}$ and $P_{\bar{\alpha}\bar{\beta}}$ where $(\alpha,\beta)=(e,\mu)$. In such a scenario, we can see in fig.~\ref{fig:prob-nue_diag} that the oscillation probabilities for Dirac and Majorana cases overlap.

%\textbf{check analytic expressions.... bla bla...}

%%%%%%%%%%%%%%%%%%%%%%%%%%%%%%
%%%%%%%%%%%%%%%%%%%%%%%%%%%%

 This implies that only a non-zero diagonal dissipative matrix can not differentiate between Dirac ($\phi = 0$) and Majorana ($\phi \ne 0$) neutrinos, which agrees with the existing results in ref.~\cite{Richter:2017toa, Capolupo:2018hrp}.

\subsection{Non-zero diagonal and off-diagonal elements in $\mathcal{D}_{mn}$:}\label{subsec:non-zero-offdiag}
In the second scenario with non-zero off-diagonal elements, we assume that the decoherence matrix $\mathcal{D}_{mn}$ in eq.~{\ref{Dmn}} has non-zero elements as in the below cases.

\begin{itemize}
    \item Case I: all diagonal elements and one off-diagonal element $\beta$ are non-zero.
    \item Case II: all diagonal elements and one off-diagonal element $\gamma$ are non-zero.
    \item Case III: all diagonal elements and one off-diagonal element $b$ are non-zero.
    \item Case IV: all diagonal and off-diagonal elements are non-zero.
\end{itemize}

%%%%%%%%%%%%%%%%%%%%%%%%%%%%%%%%%%%%
\begin{figure*}[htb!]
%\begin{center}
\centering
\includegraphics[width=0.45\linewidth]{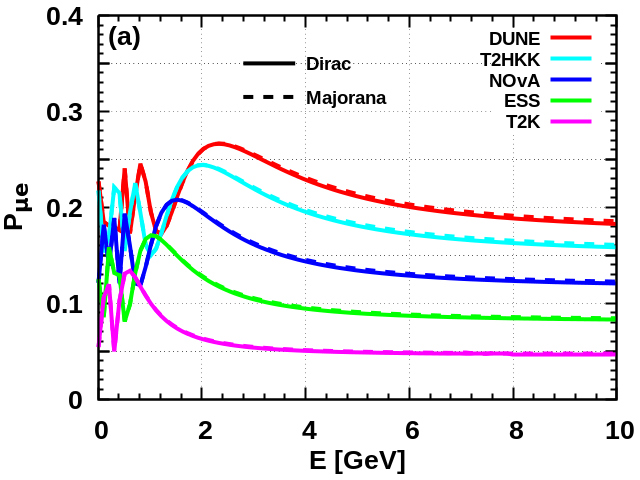}
\includegraphics[width=0.45\linewidth]{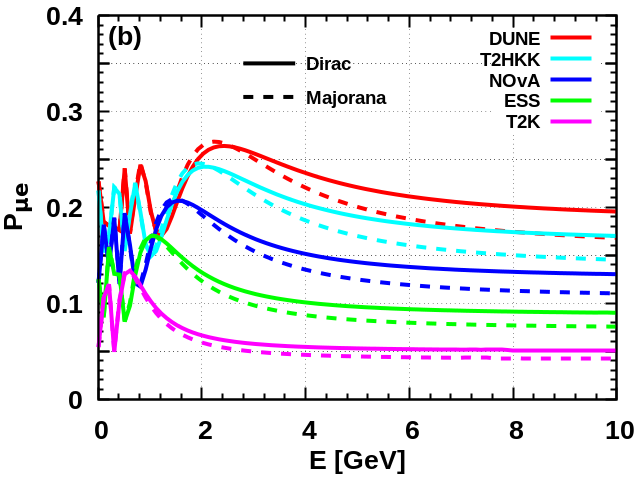}
\includegraphics[width=0.45\linewidth]{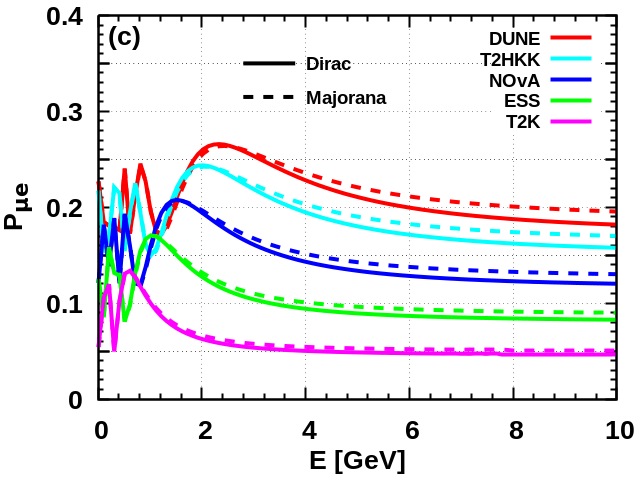}
\includegraphics[width=0.45\linewidth]{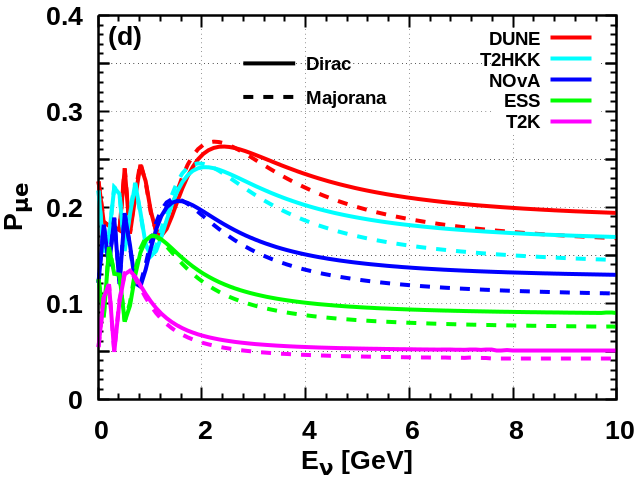}
\caption{Appearance probabilities of neutrinos with respect to energy E (assuming NH). The magenta, green, blue, cyan and red curves correspond to the baselines of 295 km, 540 km, 810 km, 1100 km and 1300 km respectively. Assumptions: (a) all diagonal and off-diagonal element $\beta$ are non-zero with $\phi = \frac{\pi}{4}$, (b) all diagonal and one off-diagonal element $\gamma$ are non-zero with $\phi = \pi$, (c) all diagonal and one off-diagonal element b are non-zero with $\phi = \frac{\pi}{2}$, and (d) all diagonal and all off-diagonal elements are non-zero with $\phi = \pi$.}
\label{fig:prob-nu}       
\end{figure*}
%%%%%%%%%%%%%%%%%%%%%%%%%%%%
\begin{figure*}[t!]
%\begin{center}
\centering
\includegraphics[width=0.45\linewidth]{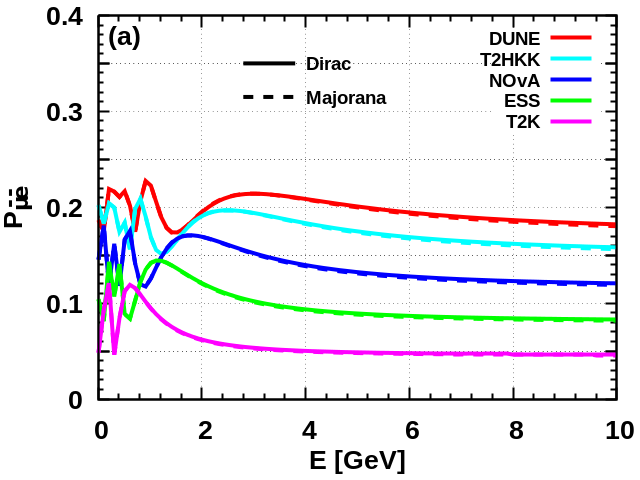}
\includegraphics[width=0.45\linewidth]{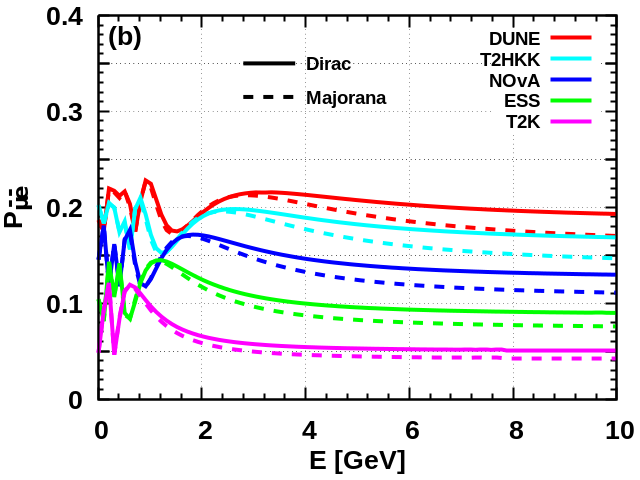}
\includegraphics[width=0.45\linewidth]{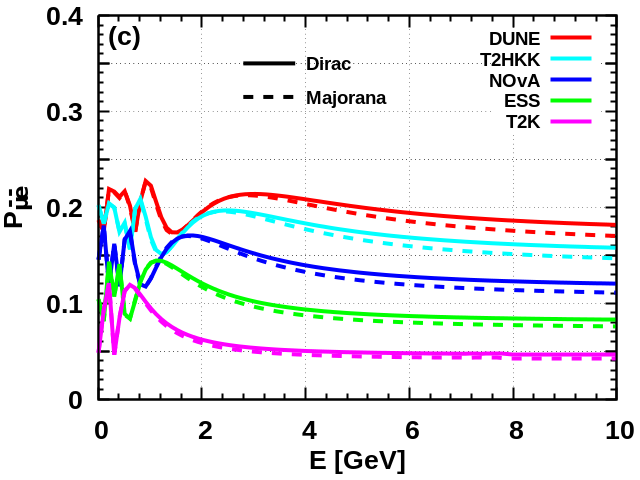}
\includegraphics[width=0.45\linewidth]{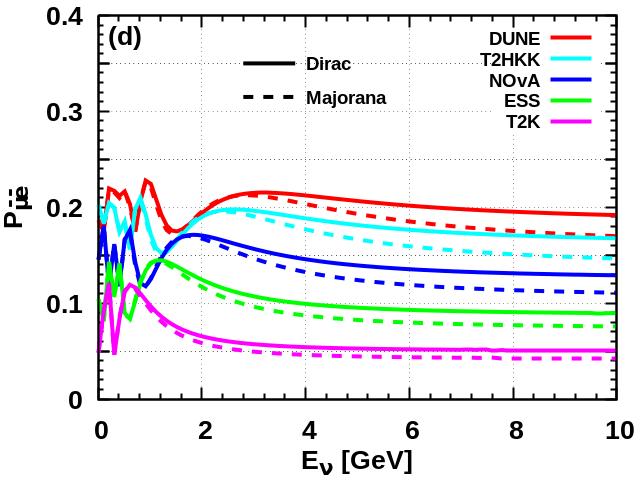}
\caption{Appearance probabilities of anti-neutrinos with respect to energy E (assuming normal mass ordering). The magenta, green, blue, cyan and red curves correspond to the probability versus energy E for the baselines 295 km, 540 km, 810 km, 1100 km and 1300 km respectively. Assumptions: (a) all diagonal and one off-diagonal element $\beta$ are non-zero with $\phi =  \frac{\pi}{4}$, (b) all diagonal and one off-diagonal element $\gamma$ are non-zero with $\phi =  \pi$, (c) all diagonal and one off-diagonal element b are non-zero with $\phi =  \frac{\pi}{2}$, and (d) all diagonal and all off-diagonal elements are non-zero with $\phi =  \pi$.}
\label{fig:prob-antinu}       
\end{figure*}
%%%%%%%%%%%%%%%%%%%%%%%%%%%%%%%%%%%%%%%%

In fig.~\ref{fig:prob-nu}, we plot the transition probabilities ($P_{\mu e}$) with respect to the neutrino beam energy E while considering the parameters given in table~\ref{table:osc-deco-params}. In fig.~\ref{fig:prob-nu}a, all diagonal and one off-diagonal element $\beta$ are non-zero and $\phi = \pi/4$; in fig.~\ref{fig:prob-nu}b, all diagonal and one off-diagonal element $\gamma$ are non-zero, $\phi = \pi$; in fig.~\ref{fig:prob-nu}c, all diagonal and one off-diagonal element b are non-zero, $\phi = \pi/2$; in fig.~\ref{fig:prob-nu}d, all elements are non-zero and $\phi = \pi$, in $\mathcal{D}_{mn}$.

The solid lines corresponding to $\phi = 0$ represent the Dirac neutrinos while the dashed lines corresponding to $\phi \neq 0$ represent the non-zero Majorana phase. The magenta, green, blue, cyan and red curves correspond to the appearance probability versus energy E for the baselines 295 km, 540 km, 810 km, 1100 km and 1300 km respectively. As can be seen in the Case I (non-zero $\beta$) in fig.~\ref{fig:prob-nu}a, the separation between the solid curves and the dashed curves is minimum across all the baselines. Whereas, in the Case II ($\gamma \neq 0$) the separation is maximum for all the baselines. Later we have observed that $\gamma \neq 0$, has consistently given rise to significant difference between Dirac and Majorana probabilities among all the four cases. This can be verified from all the subsequent figures in this work. 
At an analytic level, for $\phi = 0$ or $180\degree$, non zero terms in eq.~(\ref{nue-app prob})
 \begin{equation}\label{nue-app-prob-phi0}
 \begin{aligned}
     &P_{\mu e}(t) \\
     & = \frac{1}{2} \Bigl[1 - M_{33} \cos^2(2\theta) \mp (M_{13} + M_{31}) \sin(2 \theta) \cos(2\theta) \cos\phi \\
     & - M_{11}\sin^2(2\theta)\cos^2\phi\Bigr]~,
 \end{aligned}
 \end{equation}
'$-$' sign for $\phi = 0$ and '$+$' sign for $\phi = 180\degree$.
From the probability eq.~(\ref{nue-app-prob-phi0}) we can see that the $\cos\phi$ term is associated with $M_{13}$ and $M_{31}$. Additionally, $M_{13}$ and $M_{31}$ depend on $\mathcal{H}_{13}$ ($\gamma - \lambda \sin \phi$) and $\mathcal{H}_{31}$ ($\gamma$ + $\lambda \sin\phi$) of eq.~(\ref{H_sch}). For non-zero $\gamma$ ($\mathcal{D}_{13}$ element in eq.~(\ref{Dmn})) and $\phi = 180\degree$, one could expect a non-zero difference in the oscillation probabilities of the Dirac and Majorana neutrinos. The same reflects in the relative event rate as well as in the $\Delta \chi^2$.

In all the four figures (fig.~\ref{fig:prob-nu}), we can note that there are minimal separation between Dirac and Majorana curves for T2K baseline. 
As the baseline increases, we can see from the green (540 km), blue (810 km), cyan (1100 km) and red (1300 km) curves that the difference between solid and dashed line increases (refer to eq.~(\ref{basline-dependency})).
In the case of ESSnuSB (540 km) and T2HKK (1100 km) experiments one can see marginal differences between the Dirac (solid curve) and Majorana (dashed curve) neutrinos from the green and cyan curves at second oscillation maxima i.e. 0.2 GeV and 0.7 GeV respectively. This manifests as a decrease in the sensitivity of these experiments to differentiate between Dirac and Majorana neutrinos and is later explained in fig.~\ref{fig:event-chisq}. 
%are designed to work at the second oscillation maxima i.e. 0.2 GeV and 0.7 GeV respectively. At these energy values, one can see marginal differences between the Dirac (solid curve) and Majorana (dashed curve) neutrinos from the green and cyan curves.
While in the case of DUNE baseline one can notice that the solid and dashed red curves are well separated. Hence, DUNE could show maximum sensitivity to discriminate between Dirac and Majorana neutrinos. Nevertheless, this expectation is based on the oscillation probability plots and it may vary in case of the chi-square analyses where various systematic uncertainties of a particular experiment also play a significant role in determining the sensitivity. 

%%%%%%%%%%%%%%%%%%%%%%%%%%%% P_bar mue vs E %%%%%%%%
In fig.~\ref{fig:prob-antinu}, we present $P_{\bar{\nu}_\mu \rightarrow \bar{\nu}_e}$ with respect to anti-neutrino beam energy for all the experiments color coded as in the respective legends. The $\bar{\nu}_e$-appearance probabilities (using eq.~(\ref{antinue app prob})) for different mentioned cases are plotted w.r.t E in fig.~\ref{fig:prob-antinu}a (case I), fig.~\ref{fig:prob-antinu}b (case II) and fig.~\ref{fig:prob-antinu}c (case III). 
%In each case we plot $P_{\bar{\mu} \bar{e}}$ with E for different experiments as mentioned in the legends in each plot. 
Similar to fig.~\ref{fig:prob-nu} one can note that in all the cases the $P_{\bar{\mu} \bar{e}}$ increases as the neutrino baseline increases. We calculate the probability assuming normal hierarchy (NH) and using $\phi$ values as considered in fig.~\ref{fig:prob-nu}. 
Particularly in case II (when $\gamma$ is non-zero), we again observe the maximum difference between Dirac and Majorana phase for all the baselines. %Accordingly, the other conclusions are similar as drawn from fig~\ref{fig:prob-nu}.

%%%%%%%%%%%%%%%%%%%%%%%%%%%%%%%%%%%%%%%%%%%%%%%%%%%%%%%%%%%%%%%%

%%%%%%%%%%%%%%%%%%%%%%%%%%%% Delta-Pmue-phi %%%%%%%%%%%%%%%%%%%%%%%%%%%%%%%%%%%% 
\begin{figure*}[hbt!]
%\begin{center}
\centering
\includegraphics[width=0.45\linewidth]{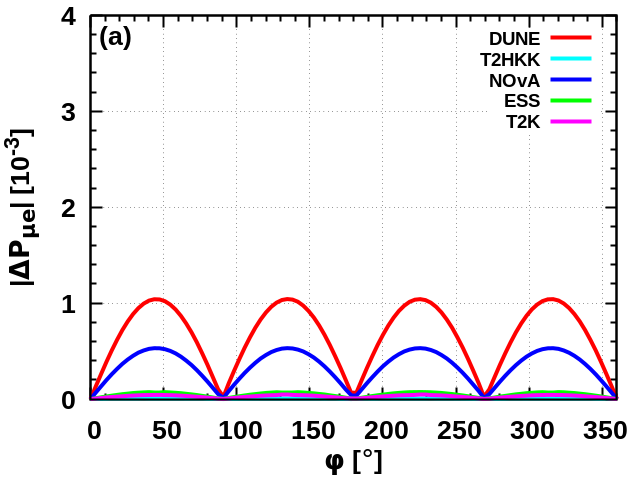}
\includegraphics[width=0.45\linewidth]{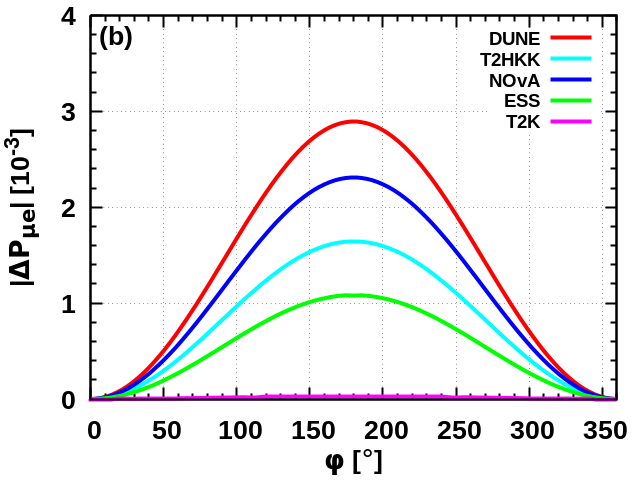}
\includegraphics[width=0.45\linewidth]{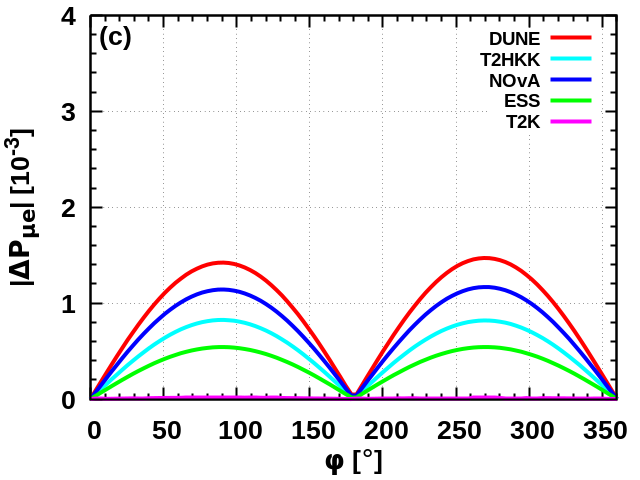}
\includegraphics[width=0.45\linewidth]{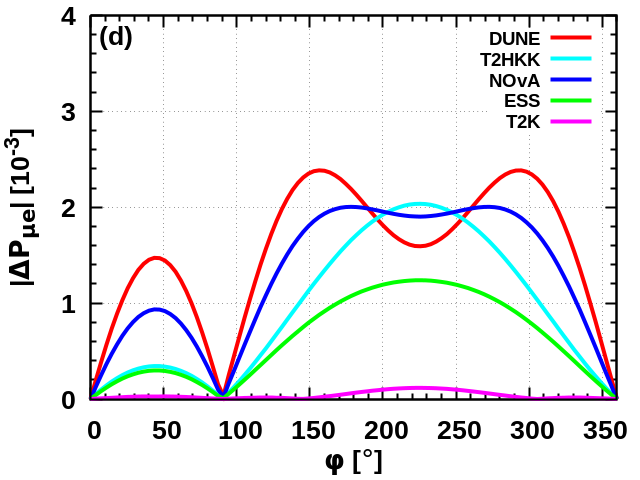}
\caption{\footnotesize $\lvert \Delta P_{\alpha \beta} \rvert$ versus $\phi$ with $(\alpha, \beta) = (e, \mu)$ using different cases in the decoherence parameters. Fig. (a), (b), (c) and (d) corresponding to case I, case II, case III and case IV respectively.}
\label{fig:delP-phi-nu}       
\end{figure*}

%%%%%%%%%%%%%%%%%%%%%%%%%%%% Delta-P_barmue-phi %%%%%%%%%%%%%%%%%%%%%%%%%%%%%%%%%%%% 
\begin{figure*}[htb!]
%\begin{center}
\centering
\includegraphics[width=0.45\linewidth]{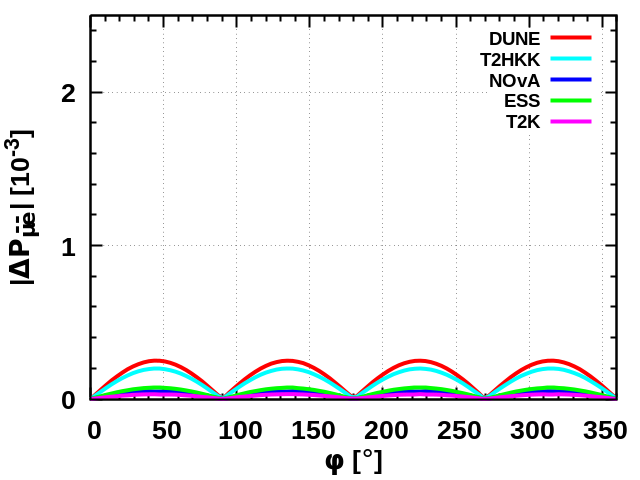}
\includegraphics[width=0.45\linewidth]{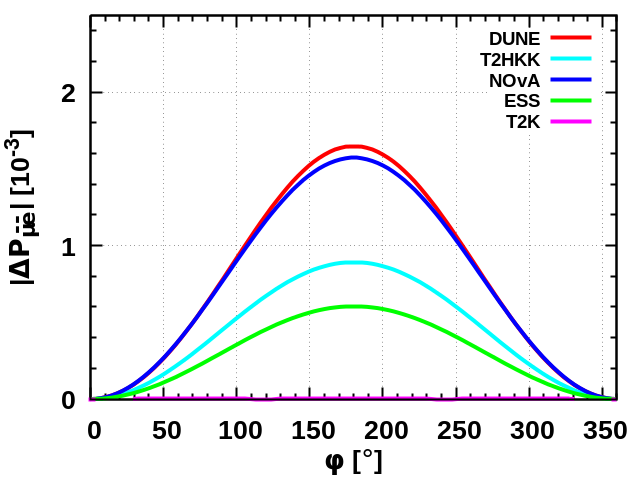}
\includegraphics[width=0.45\linewidth]{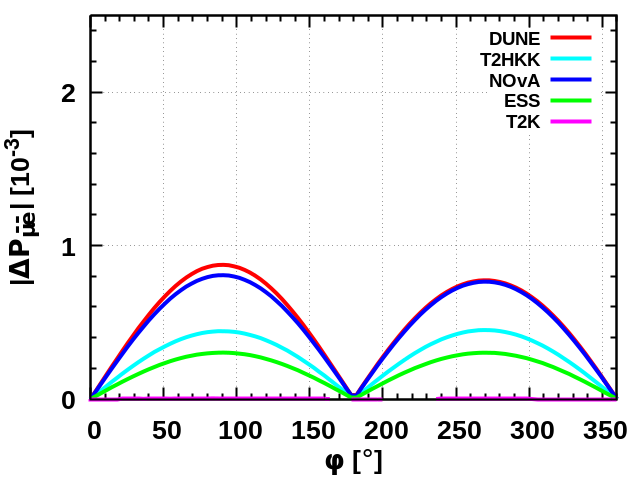}
\includegraphics[width=0.45\linewidth]{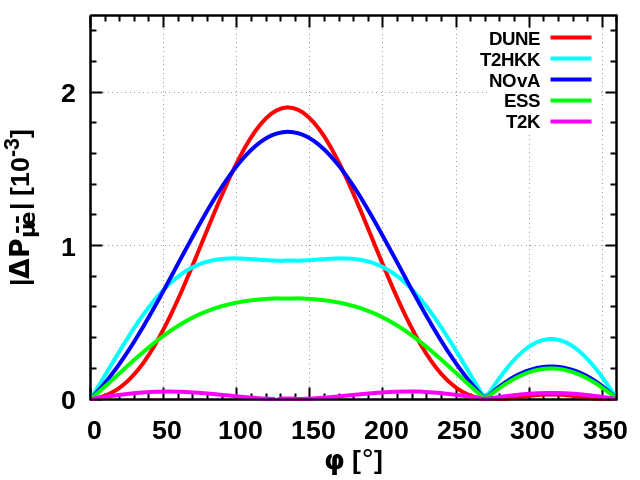} 
\caption{\footnotesize $\lvert \Delta P_{\bar{\alpha} \bar{\beta}} \rvert$ versus $\phi$ with $(\alpha, \beta) = (e, \mu)$. Fig. (a), (b), (c) and (d) corresponding to case I, case II, case III and case IV respectively.}
\label{fig:delp-antinu}       
\end{figure*}
%%%%%%%%%%%%%%%%%%%%%%%%%%%%%%%%%%%%%%%%%%%%%%%%%%%%%%%%%%%%%

 We further study the effect of decoherence on the determination of neutrino nature at the five experiments considered, by defining two quantities $\Delta P_{\alpha \beta}$ and $\Delta {P_{\bar{\alpha} \bar{\beta}}}$. 
The two quantities $\Delta P_{\alpha \beta}$ and $\Delta {P_{\bar{\alpha} \bar{\beta}}}$ in the case of neutrinos and anti-neutrinos are defined as below 
   \begin{align}\label{DelP-nu}
        \lvert \Delta P_{\alpha \beta}\rvert = \lvert P_{\alpha\beta} (Majorana) - P_{\alpha\beta} (Dirac) \rvert~,
    \end{align}
   \begin{align}\label{DelP-nubar}
        \lvert \Delta  {P_{\bar{\alpha} \bar{\beta}}}\rvert = \lvert P_{\bar{\alpha}\bar{\beta}} (Majorana) - P_{\bar{\alpha}\bar{\beta}} (Dirac)\rvert~.
    \end{align}
The first term in eq.~(\ref{DelP-nu}) and eq.~(\ref{DelP-nubar}) is obtained by assuming $\phi \subset [0,2\pi]$ while the second is obtained by taking  $\phi = 0$. When we consider $\alpha=\beta=\mu$ in eq.~(\ref{DelP-nu}) and eq.~(\ref{DelP-nubar}), one can derive that $\lvert \Delta P_{\mu \mu} \rvert = \lvert \Delta P_{\mu e} \rvert$,  $\lvert \Delta P_{\bar{\mu} \bar{\mu}} \rvert = \lvert \Delta P_{\bar{\mu} \bar{e}} \rvert$ using probability conservation for two flavour neutrino oscillations. 
%Similarly, from eq.~(\ref{DelP-nubar}), we obtain $\lvert \Delta P_{\bar{\mu} \bar{\mu}} \rvert = \lvert \Delta P_{\bar{\mu} \bar{e}} \rvert$.
Therefore, we numerically obtain two quantities $\lvert \Delta P_{\mu e} \rvert$ and $\lvert \Delta P_{\bar{\mu} \bar{e}} \rvert$ with respect to $\phi$ from eq.~(\ref{DelP-nu}) and eq.~(\ref{DelP-nubar}) to clearly quantify the discrimination between Dirac and Majorana neutrinos at the flux peak of individual experiments. Therefore, it is desirable to have a relatively larger values of $\Delta P_{\alpha \beta}$ and $\Delta {P_{\bar{\alpha} \bar{\beta}}}$ for an experiment to differentiate between Dirac and Majorana neutrinos. 

Implementing the approximations mentioned in case I, case II case III and case IV we plot $|\Delta P_{\mu e}|$ as a function of $\phi$ in figs.~\ref{fig:delP-phi-nu}a, \ref{fig:delP-phi-nu}b, \ref{fig:delP-phi-nu}c and \ref{fig:delP-phi-nu}d respectively. To obtain these figures, we fix the peak neutrino beam energy values as per the experimental specifications in table~\ref{table:exp_details} and vary the Majorana phase $\phi \subset [0,2\pi]$. We observe that $|\Delta P_{\mu e}|$ is maximum at $\phi = 45\degree$(or $135\degree$ and so on) in fig.~\ref{fig:delP-phi-nu}a, $\phi = 180\degree$ in fig.~\ref{fig:delP-phi-nu}b, $\phi = 90\degree$(or $270\degree$) in fig.~\ref{fig:delP-phi-nu}c and $\phi \sim 180\degree$ (or $270\degree$) in fig.~\ref{fig:delP-phi-nu}d. Note that the $|\Delta P_{\mu e}|$ values in fig.~\ref{fig:delP-phi-nu}a are relatively lower for all the values of $\phi$ when compared to the other two cases in fig.~\ref{fig:delP-phi-nu}b, fig.~\ref{fig:delP-phi-nu}c and fig.~\ref{fig:delP-phi-nu}d. Particularly in case II (where gamma is non-zero), we can see that relatively better differentiation between Dirac and Majorana neutrinos can be obtained at DUNE, NOvA baselines. This can be understood from the fig.~\ref{fig:prob-nu}b, where the separation between the solid and dashed lines is larger for DUNE, NOvA, T2HKK, ESS. 

In the case of ESS and T2HKK baselines one can see a reasonable separation between solid and dashed lines (i.e. dirac and majorana) around the first oscillation maxima in fig.~\ref{fig:prob-nu}b. However, this trend is not followed in fig.~\ref{fig:delP-phi-nu}b because these experiments study the second oscillation maxima (0.2 GeV for ESSnuSB and 0.7 GeV for T2HKK), and around this energy we see a negligible separation between solid and dashed lines in fig.~\ref{fig:prob-nu}b. 
However, irrespective of the cases considered, the magenta curve corresponding to the T2K baseline and the red curve corresponding to the DUNE baseline show minimum and maximum potential to determine the nature of neutrinos respectively.

%%%%%%%%%%%%%%%%%%%%%%%%%%%%%%%%%%%%%%%%%%%%%%%
In fig.~\ref{fig:delp-antinu} we show $\lvert \Delta P_{\bar{\mu} \bar{e}} \rvert$ w.r.t $\phi$ for mentioned cases. We set the energy value at the flux peak of the individual experiments and calculate $\lvert \Delta P_{\bar{\mu} \bar{e}} \rvert$ as in eq.~(\ref{DelP-nubar}). Analyzing fig.~\ref{fig:delp-antinu} one can draw similar conclusions as in fig.~\ref{fig:delP-phi-nu}~.
%%%%%%%%%%%%%%%%%%%%%%%%%%%%%%%%%%%%%%%%%%%%%%%%%%%%%%%%%%%%%%%%%

\section{Event rates and sensitivity}\label{sec:event-sensitivity}
\begin{figure*}[htb!]
%\begin{center}
\centering
\includegraphics[width=0.4\linewidth,height=4.4cm]{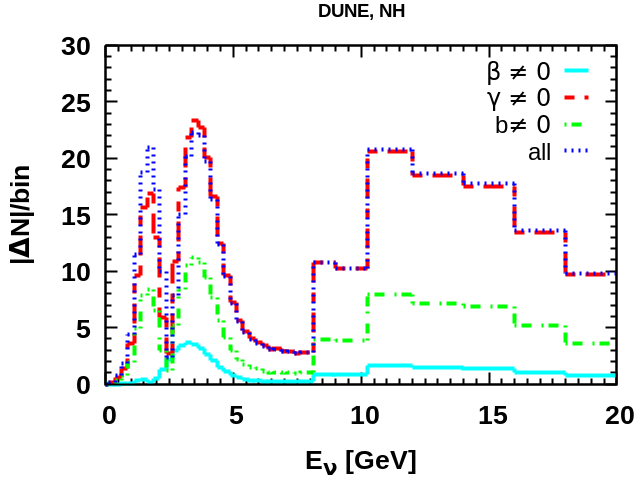}
\includegraphics[width=0.4\linewidth,height=4.4cm]{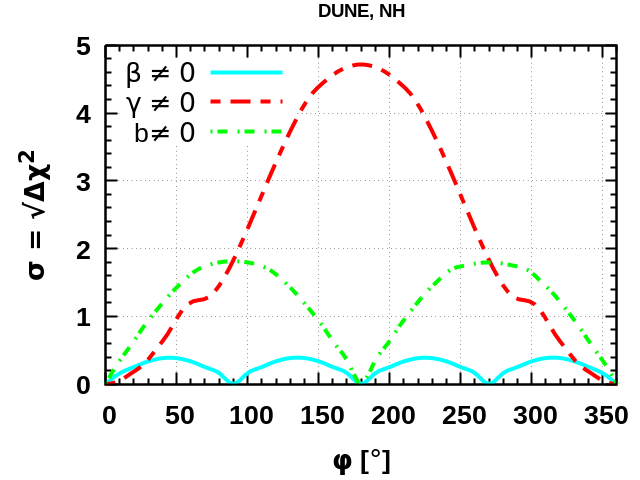}
\includegraphics[width=0.4\linewidth,height=4.4cm]{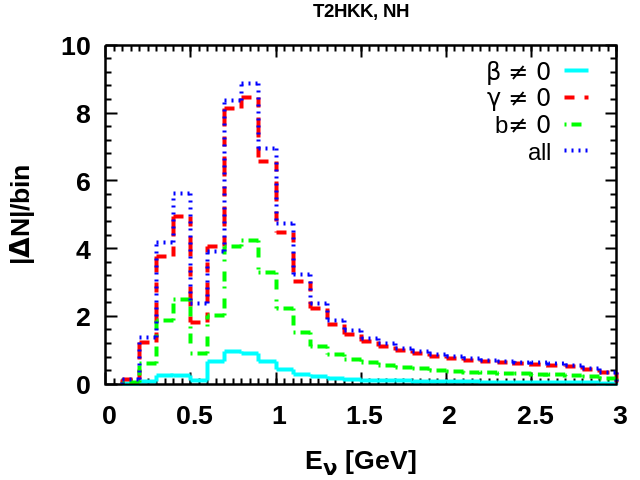}
\includegraphics[width=0.4\linewidth,height=4.4cm]{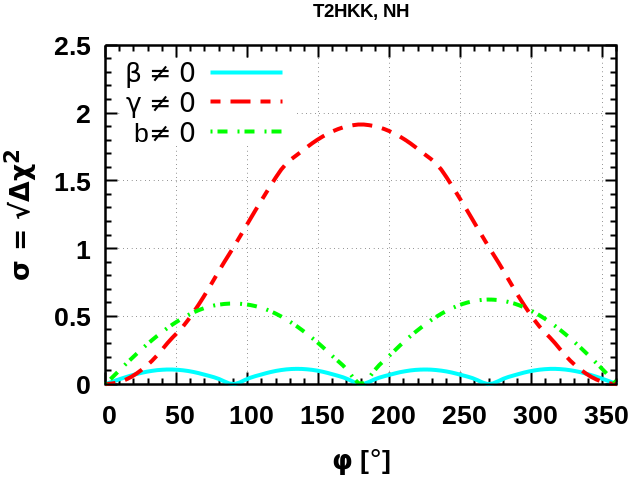}
\includegraphics[width=0.4\linewidth,height=4.4cm]{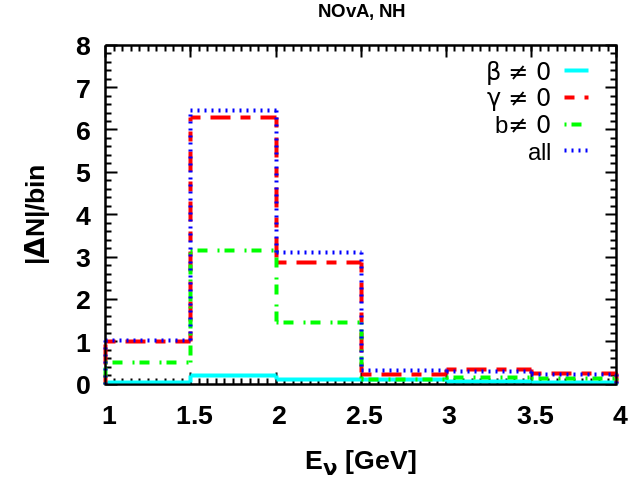}
\includegraphics[width=0.4\linewidth,height=4.4cm]{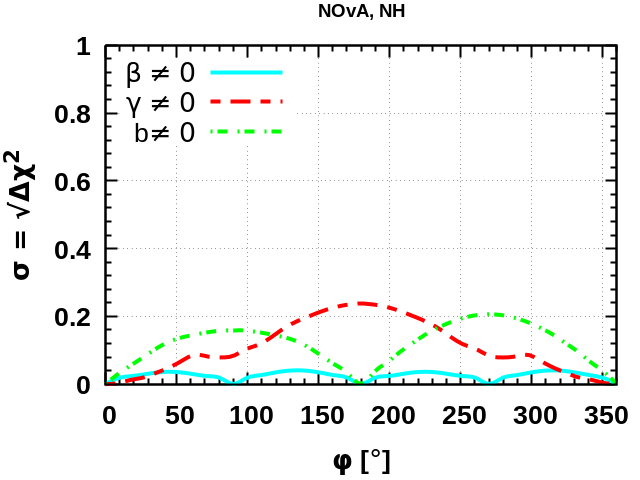}
\includegraphics[width=0.4\linewidth,height=4.4cm]{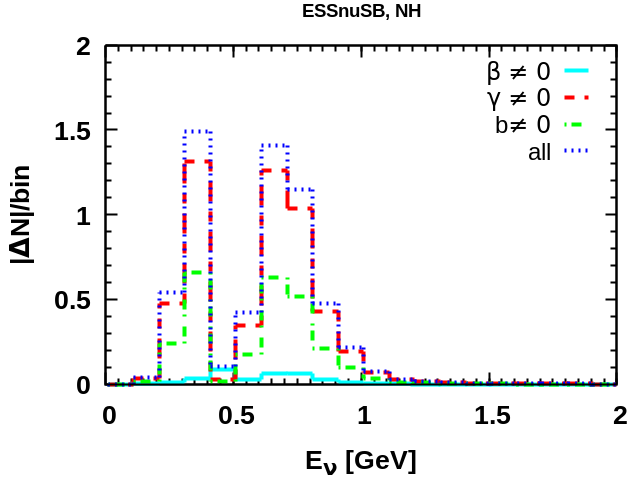}
\includegraphics[width=0.4\linewidth,height=4.4cm]{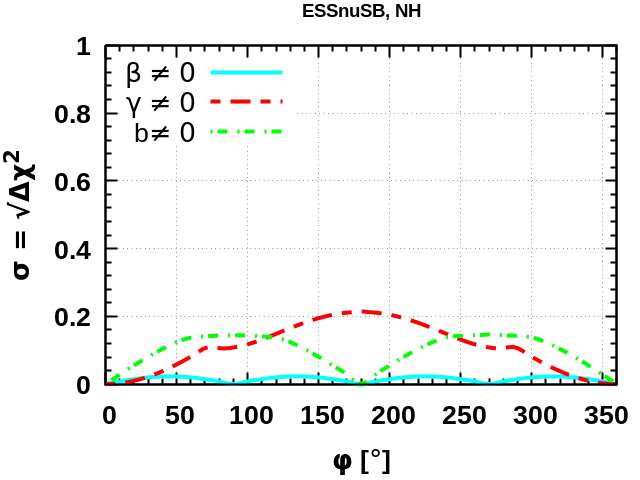}
\caption{\footnotesize In the left panel, relative $\nu_e$ appearance events per bin w.r.t $E_\nu$. In the right panel, $\sigma$ versus Majorana phase $\phi$. In the first, second, third and fourth rows corresponding to the results for DUNE, T2HKK, NOvA and ESSnuSB respectively.}
\label{fig:event-chisq}       
\end{figure*}

In this section, we present the the event rates (fig.~\ref{fig:event-chisq}) to demonstrate the sensitivity of different experiments in determining the neutrino nature. Based on the analyses in the previous section (at probability level), we note that T2K experiment shows negligible $\Delta P$ at any $\phi$ both for neutrino and antineutrino appearance channel. Therefore, we exclude T2K experiment in the present section. We only discuss the event rates and sensitivities of DUNE (first row), T2HKK (second row), NOvA (third row) and ESSnuSB (fourth row).

%%%%%%%%%%%%%%%%%%%%%%%%%%%%%%%%%%%%%%%%%
%%%%%%%%%%%%%%%%%%%%%%%%%%%%%%%%%%%%%%%%%%%%%%%%%%%%%%%%%%%%%

 In the left panel of fig.~\ref{fig:event-chisq} we show the relative event rate per bin defined as, 
 \begin{equation}
     \lvert\Delta N \rvert = \lvert N(Majorana) - N(Dirac) \rvert~,
 \end{equation}
 as a function of neutrino energy.  In each plot the cyan, red, green and blue curves represent the $\Delta N$ corresponding to case I, case II, case III and case IV respectively.
The cyan curve (case I) shows lowest $\Delta N$, where as the red (case II) and blue curves (case IV) show highest value of $\Delta N$ for all the experiments.
% From individual experiment we can see the relative event rate very low for case-I, comparatively higher for case-II and moderate for case-III.
We observe that the $\Delta N$ corresponding to case II and case IV are similar. We can explain this from the corresponding probability and relative probability plots in the previous section. 
%Note that, the cases with non-zero $\gamma$ show maximum event differences in Dirac and Majorana phase. 

 In the right panel of fig.~\ref{fig:event-chisq} we display the $\sigma (= \sqrt{\Delta \chi^2})$ as a function of Majorana phase $\phi$. We define $\Delta \chi^2$ as,
 \begin{equation}
     \Delta \chi^2 = min(\chi^2_{true} (\phi \neq 0) - \chi^2_{test} (\phi = 0))~,
 \end{equation}
and marginalize over $\theta$, $\Delta m^2$ and decoherence parameters to obtain minimum $\Delta \chi^2$. Since, $\Delta N$ are similar for case II and case IV, these two cases give similar significance. Considering this we would like to discuss the $\chi^2$ analysis for case I (cyan curve), case II (red curve) and case III (green curve) 
%Considering above equation we discuss the $\chi^2$ analysis for case I (cyan curve), case II (red curve) and case III (green curve)
to illustrate the sensitivity of various experiments to determine the neutrino nature. 

We observe that case I parades very low sensitivity, case III exhibits moderate sensitivity and case II displays comparatively higher sensitivity in all the experiments. This trend is similar to the corresponding $\Delta P$ (fig.~\ref{fig:delP-phi-nu},~\ref{fig:delp-antinu}) and $\Delta N$ (fig.~\ref{fig:event-chisq}). Interestingly, case II ($\gamma \neq 0$) offers maximum sensitivity for the $\phi \sim \pi$ and minimum sensitivity at $\phi = 0,360\degree$. This behavior can be explained by the presence of $\cos\phi$ in the appearance probability expression for neutrinos and antineutrinos in eq.~(\ref{nue-app prob}) and eq.~(\ref{antinue app prob}). From the probability equation we can see that the $\cos\phi$ term is associated with $M_{13}$ and $M_{31}$. Additionally, $M_{13}$ and $M_{31}$ depend on $\mathcal{H}_{13}$ ($\gamma - \lambda \sin \phi$) and $\mathcal{H}_{31}$ ($\gamma$ + $\lambda \sin\phi$) of eq.~(\ref{H_sch}). For non-zero $\gamma$ ($\mathcal{D}_{13}$ element in eq.~(\ref{Dmn})) and $\phi = 180\degree$, one could expect a non-zero difference in the oscillation probabilities of the Dirac and Majorana neutrinos. Which reflects in the relative event rate as well as in the $\Delta \chi^2$.

However, DUNE shows sensitivity more than $\sim 3\sigma$ CL specially for the $\gamma \neq 0$ (case II) and $\phi \sim 180\degree$ as it is a broad beam experiment. Case I and case III show sensitivity less than $2\sigma$ for all values of $\phi$. The case II corresponding to T2HKK experiment presents sensitivity $\sim 2\sigma$ (at $\phi \sim 180\degree$) and rest two cases show $< 1\sigma$ sensitivity for all $\phi$. Although T2HKK has high statistics but because of the flux peak at second oscillation maxima, we observe comparatively lesser event rate, which leads to a low $\Delta \chi^2$ for this experiment. On the other hand, NOvA and ESSnuSB experiments show marginal sensitivity to differentiate Dirac and Majorana neutrinos for all values of $\phi$ and for all the cases because of relatively low statistics.

%%%%%%%%%%%%%%%%%%%%%%%%%%%%%%%%%%%%%%%%%%%%%%%%%%%%%%%%
\section{Conclusions}\label{sec:conclusion}
We analyse the two flavor neutrino oscillations in matter in a dissipative environment. Firstly, for simplicity, we consider the case where we have non-zero diagonal elements and zero off-diagonal elements in the decoherence matrix. Whereas in the second case, we assume non-zero diagonal and non-zero off-diagonal elements. We see that the transition probabilities in the former case do not depend on Majorana phase $\phi$, on the other hand in the later case the transition probabilities depend explicitly on the Majorana phase $\phi$.
%We discuss the modified transition probabilities relevant to long baseline neutrino oscillation experiments and study their dependency on the Majorana phase $\phi$ and Dirac neutrinos by assuming $\phi = 0$.
%On the other hand, we also obtain the oscillation probabilities corresponding to the Dirac neutrinos by assuming $\phi = 0$.
We further study the oscillations of Dirac and Majorana neutrinos at L (baseline) and E (neutrino beam energy) values corresponding to five long-baseline neutrino oscillation experiments T2K, NOvA, ESSnuSB, T2HKK and DUNE.

We observe that the experiments focusing on first oscillation maxima (DUNE, NOvA) with relatively longer baseline are more effective in distinguishing between Dirac and Majorana neutrinos. Whereas, the experiments like ESSnuSB, T2HKK which study the second oscillation maxima show relatively poor sensitivity to differentiate between Dirac and Majorana neutrinos. 
Therefore one can probe the nature of neutrinos at these experiments when the neutrino subsystem interacts with the environment, leading to a dissipative matrix containing at least one non-zero off-diagonal element along with the diagonal elements. Specifically, cases with $\gamma \neq 0$ (in $\mathcal{D}_{mn}$) show maximum difference in $|\Delta P|$ compared to the other cases. In particular, case II ($\gamma \neq 0$) at $\phi \sim 180\degree$ shows large sensitivity compared to other cases. In conclusion, among all the experiments DUNE shows maximum sensitivity to discriminate between Dirac and Majorana neutrinos.

\bibliographystyle{elsarticle-num}
\bibliography{thebibliography}

\end{document}